\documentclass[a4paper,fleqn,usenatbib]{mnras}

\usepackage[T1]{fontenc}
\usepackage{ae,aecompl}

\usepackage{graphicx}	
\usepackage{amsmath}	
\usepackage{amssymb}	
\usepackage{multicol}        
\usepackage{bm}		
\usepackage{color}
\usepackage{xcolor}
\usepackage{soul}         

\title[Spots on hot stars]
{Evidence for spots on hot stars suggests major revision of stellar
physics}
\author[L.A. Balona]{L. A. Balona\\
South African Astronomical Observatory, P.O. Box 9, Observatory, Cape
Town, South Africa}

\begin{document}

\date{Accepted .... Received ...}

\pagerange{\pageref{firstpage}--\pageref{lastpage}} \pubyear{2011}

\maketitle

\label{firstpage}

\begin{abstract}
It has long been thought that starspots are not present in the A and
B stars because magnetic fields cannot be generated in stars with 
radiative envelopes.  Space observations show that a considerable 
fraction of these stars vary in light with periods consistent with the expected
rotation periods.  Here we show that the photometric periods are the same
as the rotation periods and that starspots are the likely cause for the
light variations.  This discovery has wide-ranging implications and suggests 
that a major revision of the physics of hot stellar envelopes may be required.
\end{abstract}

\begin{keywords}
stars:early type; stars: rotation; stars:starspots
\end{keywords}

\section{Introduction}

It is accepted that the outer envelopes of main sequence stars with effective 
temperatures hotter than about 7000\,K are in radiative equilibrium.  The lack 
of convection in the outer layers precludes the operation of the dynamo 
mechanism which is believed to be necessary to generate surface magnetic 
fields \citep{Charbonneau2014}.    Indeed, measurements in two bright A stars, 
Vega and Sirius, indicate global magnetic fields of less than 1\,G, which is 
weaker than that of the Sun \citep{Petit2011}.  For this reason, photospheric 
activity such as starspots and flares, are not expected in A and B stars.

This picture of quiescent radiative envelopes, in which diffusion and 
gravitational settling can proceed relatively undisturbed, has been 
successful in accounting for peculiar A and B stars.  This process,
operating in the absence of a magnetic field and in the absence of mixing by
convection or rotational circulation, is generally accepted as the
explanation for the metallic-lined Am stars \citep{Michaud1976}. The same
process operating in the presence of a strong global magnetic field, is
thought to be responsible for the patches of anomalous abundances in the
chemically peculiar Ap and Bp stars \citep{Michaud1976}.  The kilogauss global 
magnetic fields in Ap and Bp stars are presumed to be of fossil 
origin \citep{Braithwaite2004}.  

For cool stars with convective envelopes, a magnetic field in conjunction with 
a stellar wind exerts a torque on the ejected matter, resulting in a steady 
loss of angular momentum.  On the other hand, hotter stars with radiative 
envelopes do not experience loss of angular momentum in this way.  This 
explains the steep increase of rotation rate between main-sequence stars with 
convective and radiative envelopes.  Should it be found that spots are present 
in stars with radiative envelopes, just as they are in cool stars with 
convective envelopes, the ideas described above will most probably require 
revision.

Photometric observations of very high precision from space, particularly by 
the {\it Kepler} and {\it TESS} missions, have gradually revealed a picture
which is at odds with our current understanding of stars with radiative
envelopes. Pulsational driving in the $\delta$~Scuti stars, which have
effective temperatures in the range 6500--9000\,K, is thought to be a 
result of the $\kappa$ opacity mechanism operating in the HeII ionization zone.
Models predict pulsation modes with frequencies greater than about 
6\,d$^{-1}$. However, the first {\it Kepler} observations revealed that a 
large fraction of $\delta$~Scuti stars also pulsate in numerous low frequency 
modes \citep{Grigahcene2010}, in conflict with model predictions. It is now 
known that at least 98\,percent of $\delta$~Scuti stars contain low 
frequencies \citep{Balona2018c}.

\begin{table*}
\begin{center}
\caption{Extract from the on-line catalogue.  The star name
and variability type classification is followed by the $V$ magnitude,
effective temperature and luminosity.  The projected rotational velocity,
presumed rotational frequency, $\nu$, and its error is followed by the
rotational amplitude, $A_1$, and its error in parts per thousand (ppt).  
$A_2$ is a least-squares fit to the first harmonic, $2\nu$, followed by its 
error.  The signal-to-noise ratio (S/N) of $A_1$ is given as well as the
spectral type when available.}
\label{data}
\resizebox{18cm}{!}{
\begin{tabular}{llrrrrrrrrrrrl}
\hline
\multicolumn{1}{c}{Name}   & 
\multicolumn{1}{c}{Var Type}   & 
\multicolumn{1}{c}{$V$}   & 
\multicolumn{1}{c}{$T_{\rm eff}$}   & 
\multicolumn{1}{c}{$\log \tfrac{L}{L_\odot}$}   & 
\multicolumn{1}{c}{$v\sin\,i$}  &
\multicolumn{1}{c}{$\nu$}   & 
\multicolumn{1}{c}{e\_$\nu$}   & 
\multicolumn{1}{c}{$A_1$}   & 
\multicolumn{1}{c}{e\_$A_1$}   & 
\multicolumn{1}{c}{$A_2$}   & 
\multicolumn{1}{c}{e\_$A_2$}   & 
\multicolumn{1}{c}{S/N}   & 
\multicolumn{1}{c}{Sp Type}     \\
\multicolumn{1}{c}{}   & 
\multicolumn{1}{c}{}   & 
\multicolumn{1}{c}{mag}   & 
\multicolumn{1}{c}{K}   & 
\multicolumn{1}{c}{dex}   & 
\multicolumn{1}{c}{km\,s$^{-1}$}  &
\multicolumn{1}{c}{d$^{-1}$}   & 
\multicolumn{1}{c}{d$^{-1}$}   & 
\multicolumn{1}{c}{ppt}   & 
\multicolumn{1}{c}{ppt}   & 
\multicolumn{1}{c}{ppt}   & 
\multicolumn{1}{c}{ppt}   & 
\multicolumn{1}{c}{}   & 
\multicolumn{1}{c}{}     \\
\hline
KIC 3331147   & ROT+FLARE &       10.060 &  7000 & 0.71 &  63.0 & 1.410508 & 0.000003 &  4.747 & 0.037 &  0.104 & 0.037 &  24.1 & F0.5V  \\
KIC 4570326   & DSCT+ROT  &        9.760 &  7000 & 1.50 &  80.0 & 0.892275 & 0.000015 &  2.677 & 0.109 &  1.425 & 0.109 &  92.5 & F1V    \\             
KIC 6128236   & ROT       &        8.888 &  7000 & 1.51 & 106.0 & 0.854585 & 0.000008 &  0.048 & 0.001 &  0.001 & 0.001 &  30.5 &        \\             
KIC 9651374   & ROT       &       11.683 &  7000 & 0.88 &       & 0.362966 & 0.000003 &  0.184 & 0.001 &  0.023 & 0.001 &  62.1 &        \\             
TIC 100101337 & ROT       &        7.276 &  7000 & 1.02 &       & 1.440783 & 0.000293 &  0.072 & 0.002 &  0.062 & 0.002 &   9.1 & F0IV   \\             
TIC 119983704 & ROT       &        8.550 &  7000 & 0.83 &       & 1.573979 & 0.000185 &  2.960 & 0.019 &  0.211 & 0.019 &  11.2 & F0IV/V \\             
\hline                        
\end{tabular}
}
\end{center}
\end{table*}

The huge disparity in pulsation frequency distributions among $\delta$~Scuti 
stars with the same effective temperature and luminosity and the fact that 
less than half of the stars in the instability strip actually pulsate 
\citep{Balona2018c} also present serious challenges.  Another problem is the 
presence of $\delta$~Scuti pulsations in stars which are much hotter than 
predicted (the Maia variables: \citealt{Mowlavi2013,Balona2016c}).  

The problem involving stellar pulsation among the A stars is a severe 
challenge, but this is further compounded by {\it Kepler} observations which 
suggest that a large fraction of A and B stars vary with periods which are 
consistent with their rotation periods, suggesting the presence of starspots
\citep{Balona2013c, Balona2016a, Balona2019a}.   Until then, starspots 
were believed to be present only in cool stars with convective envelopes.

The advent of {\it TESS} has greatly increased the sample of A and B stars
in which possible rotational modulation can be detected.  In this paper we use 
data from the {\it Kepler}, {\it K2} and {\it TESS} missions to show that the 
photometric period is indistinguishable from the rotation period.  
The implication is that spots are present in A and B stars and that the 
current understanding of stars with radiative atmospheres may need to be 
revised.

\section{Data}

The data used in this study comprises light curves from the full four-year 
{\it Kepler} mission, from the {\it K2} mission and from sectors 1--13 of the 
{\it TESS} mission.  Corrected data using pre-search data conditioning (PDC) 
were used for {\it Kepler} and {\it TESS}.  For {\it K2} data, light curves 
corrected by the method described in \citet{Vanderburg2014} were used.

Each star was assigned, where appropriate, a variability type by visual 
inspection of the periodogram and light curve with the assistance of the 
spectral type or effective temperature.  As far as possible, the classification
scheme used in the {\it General Catalogue of Variable Stars} \citep{Samus2009} 
was followed.  For example, an A-type star with frequency peaks of 5\,d$^{-1}$ or 
higher is a $\delta$~Scuti variable, whereas if it is an early B star it would 
be classified as a $\beta$~Cephei variable.  Stars with multiple frequency peaks 
lower than this value are classified as $\gamma$~Doradus or SPB respectively.  

Many stars do not fit in this scheme.  For example, a large fraction of A
and B stars have a single peak or a peak and its harmonic at a low frequency
(i.e. less than about 4\,d$^{-1}$). Since there is no known pulsation
mechanism which can explain such frequencies in early A stars, and since the
frequencies are consistent with the expected rotational frequencies, these
were given a preliminary classification ROT (rotational variable).  Because
of the very low amplitudes, a classification in terms of binarity appeared
unlikely.  When sufficient numbers of these ROT variables became available,
various statistical tests increasingly supported the idea that these stars
are indeed rotational variables \citep{Balona2011e, Balona2013c,
Balona2015c, Balona2016a, Balona2017a}.

The advent of the {\it TESS} mission has greatly increased the number of stars 
with effective temperatures greater than 7000\,K.  There are now 2861 
of these stars with known photometric periods.  This allows a more rigorous
test of the rotational modulation hypothesis than is possible using only 
{\it Kepler} and {\it K2} data.  A catalogue of the rotation frequencies, 
amplitudes and other information is available in electronic form.  An extract 
of the catalogue is shown in Table\,\ref{data}.
 
The projected rotational velocity, $v\sin\,i$, is mainly from 
\citet{Glebocki2005}. The effective temperatures, $T_{\rm eff}$, are mostly 
from the {\it Kepler Input Catalogue} \citep{Brown2011a} corrected for A stars 
in accordance with the recipe in \citet{Balona2015d}.   For the B stars, the 
effective temperature is mostly from the literature.  The luminosities are 
determined using {\it Gaia}~DR2 parallaxes \citep{Gaia2016,Gaia2018} using
bolometric corrections from \citet{Pecaut2013} and a 3D interstellar 
extinction model by \citet{Gontcharov2017}.  

The median rotational amplitude is 93\,ppm, which is far too small to be 
detected from the ground.   In about half of the stars, the harmonic of the 
presumed rotation period is  visible in the periodogram.  For this reason,
the table lists the amplitude of the first harmonic derived from a
least-squares fit. The signal-to-noise ratio, S/N, of the rotation frequency is 
the ratio of the peak height to the mean surrounding noise level in the 
periodogram.  Its median value is S/N=14.  Finally, the spectral type, 
mostly from the catalogue of \citet{Skiff2014}, is listed.  

There are several groups of stars of particular interest in the catalogue,
including 367 $\delta$~Sct or $\gamma$~Dor stars.  In these stars, a 
prominent peak and its harmonic are visible in the periodogram and assumed
to be a result of rotational modulation.  Flares are seen in 51 A or B
stars.  There are 214 stars with a peculiar pattern consisting of a 
sharp peak in the periodogram flanked by a broad peak at slightly lower 
frequency.  These have been assigned the variability type ROTD
\citep{Balona2013c, Balona2015c}.  The classification of a star as a 
rotational variable is, of course, subjective.  The distinction between
rotation and binarity is based on amplitude variability, the shape of the
light curve and low amplitude.

\section{Results}

From $T_{\rm eff}$ and $\log L/L_\odot$ (estimated from the {\it Gaia} DR2 
parallax), the stellar radius can be calculated.  Using this radius and the
presumed rotation period from the photometry, the equatorial rotational
velocity, $v$, can be determined.  If the photometric period is the rotation 
period, then there should be a relationship between $v$ and the projected 
rotational velocity, $v\sin\,i$, derived from spectroscopy.   

Because of the unknown inclination angle, $i$, such a comparison needs to be 
made statistically using a large sample of stars.  Unfortunately $v \sin\,i$ 
measurements are available for very few A and B stars in {\it Kepler} and 
{\it K2}.  A comparison using 30 stars with $T_{\rm eff}$ in the range 
8300-12000\,K supports the identification of the photometric period with the 
rotation period \citep{Balona2017a} .  A similar study using {\it TESS} 
observations of B stars comes to the same conclusion \citep{Balona2019a}. 
Both these studies involve small numbers of stars.

\begin{table}
\begin{center}
\caption{The number of stars with known photometric periods, $n$, and the total 
number of stars, $N$, in the given effective temperature range, $T_{\rm eff}$, 
is given.  The number of stars with both photometric periods and $v\sin\,i$ 
measurements is $n_{v\sin\,i}$. The number of stars used in the 
$v\sin\,i$ frequency distribution is $N_{v\sin\,i}$.}
\label{tab}
\begin{tabular}{rrrrr}
\hline
\multicolumn{1}{c}{$T_{\rm eff}$}   & 
\multicolumn{1}{c}{$n$}   & 
\multicolumn{1}{c}{$N$}   & 
\multicolumn{1}{c}{$n_{v\sin\,i}$}  &
\multicolumn{1}{c}{$N_{v\sin\,i}$}     \\
\hline
  7000--8000 &  1609 &  5852 &   81 &  455 \\
 8000--10000 &  1035 &  2675 &  113 &  372 \\
10000--15000 &   217 &   671 &   80 &  582 \\
\hline                        
\end{tabular}
\end{center}
\end{table}

\begin{figure}
\centering
\includegraphics[]{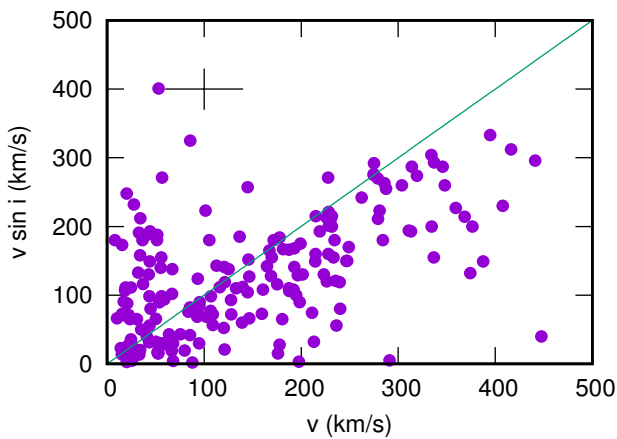}
\caption{The projected rotation velocity, $v\sin\,i$, as a function of the
equatorial rotation velocity, $v$, estimated from the photometric period
for stars with effective temperatures $T_{\rm eff} > 8000$\,K.  The straight
line is $\sin\,i~=~1$. The 1-$\sigma$ error bars are shown on the top left.}
\label{vsini}
\end{figure}

{\it TESS} observations have revealed the rotation frequencies of many more A 
and B stars for which $v\sin\,i$ measurements are available.  Table\,\ref{tab} 
shows that there are 274 stars with both rotation frequencies and $v\sin\,i$.  
In Fig.\,\ref{vsini} a comparison between $v\sin\,i$ and the 
photometric $v$ is made for 193 stars with $T_{\rm eff} > 8000$\,K.  The 
temperature limit was chosen to minimise contamination by stars in which 
surface convection may still be present.  Since $v\sin\,i$ cannot exceed $v$, 
the points are expected to lie below the $\sin\,i~=~1$ line in the figure.   

The increase in numbers of stars above the $\sin\,i~=~1$ line at low rotation
rates is to be expected.  Measurements of $v\sin\,i$ are mostly obtained by 
measuring the widths of selected spectral lines and applying a calibration to 
convert to $v \sin\,i$.  The calibration is obtained using high-dispersion 
spectra and modelling line profiles.  The error in $v \sin\,i$ is 
approximately constant at all values of $v\sin\,i$.  Analysis of the rotational 
velocity catalogue of \citet{Glebocki2005} gives a standard deviation of 
$v\sin\,i$ of about 30\,km\,s$^{-1}$.  This means that at low values of 
$v\sin\,i$, the error is comparable to, or larger, than $v\sin\,i$ itself and 
many points will lie above the $\sin\,i~=~1$ line.  This is further compounded
by the fact that $v \sin\,i$ is constrained to be greater than or equal to
zero, although measurements of line width may lead to negative values of
$v\sin\,i$ on applying the calibration.  As the rotation rate tends to zero,
$v \sin\,i$ measurements of ever increasing precision are required to
ensure that $v\sin\,i$ stays below the $\sin\,i~=~1$ line.  This is clearly
not possible to attain.

Another consideration is that low photometric frequencies are more difficult 
to measure because a longer time span is required to resolve the frequency 
peak in the periodogram.  Thus instrumental drift becomes important.  This is 
further compounded by the fact that the periodogram noise increases sharply 
towards zero frequency.  Thus the uncertainty in $v$ increases as $v$ 
approaches zero, contributing to moving the point above the $\sin\,i~=~1$ line.
The standard deviation for $v$ is about 40\,km\,s$^{-1}$ as estimated from
the errors in $T_{\rm eff}$ and $\log L/L_\odot$.  

Finally, of course, the possibility exists that some of the stars classified
as rotational variables are, in fact, binaries.  The distinction between
light variability due to binarity effects and rotation was made using two
principles.  Firstly, stars with low frequencies and amplitudes higher than
a few parts per thousand are assumed to be binaries.  However, if there is
an indication of amplitude variability (seen as broad peaks in the
periodogram or amplitude changes in the light curve), then the star is
classified as a rotational variable.  The distinction becomes increasingly
difficult at low frequencies because of instrumental drift and the overall
increase in periodogram noise towards low frequencies.  Classification
becomes more uncertain at low frequencies.

Since most of the stars would be observed at high angles of inclination, a 
trend between $v\sin\,i$ and $v$ is expected, as seen in Fig.\,\ref{vsini}.
If a linear relationship between $v \sin\,i$ and $v$ is assumed, the probability
of the observed correlation occurring by chance, as measured by Student's t 
test is less than $10^{-7}$.  In actual fact the correlation is not exact
because of the $\sin\,i$ factor.  If it was possible to take this into
account, the correlation will be even higher.  The stated probability is 
thus an upper limit, clearly establishing that a physical effect exists.  
Whatever is responsible for the photometric variability must be intimately 
connected with rotation.

It should be noted that there is a inherent difference in the way the
rotation rate is measured using $v\sin i$ and the photometric frequency.  If
the photometric variability is a result of starspots, then there will be a
lack of stars with low inclinations because the rotational modulation
amplitude tends to zero as $i$ tends to zero.  The underlying distribution
of stars measured by the two methods is not the same, which means that until
we know the distribution of spot sizes and locations, in addition to the
distribution of the angles of inclination, there is no possibility
of a rigorous statistical test to compare the photometric $v$ with 
the spectroscopic $v\sin i$.   

The equatorial rotation velocity distribution, i.e. the relative number of 
stars within a particular range of equatorial rotation velocity, is an
important quantity which provides information on the physics of stellar
rotation.  The distribution is expected to vary as a function of effective
temperature and gravity due to various factors such as mass loss and
evolutionary state.  It is obtained from a large number of $v\sin\,i$ 
measurements of stars within a limited $T_{\rm eff}$ and evolutionary state by 
deconvolving $v \sin\,i$ assuming random orientation of the axes of rotation.
The process requires the inversion of an integral equation
\citep{Chandrasekhar1950}.

The photometric rotation frequencies, in combination with radii obtained
from {\it Gaia} DR2 parallaxes, allows the equatorial rotation velocity 
distribution to be obtained without the need for deconvolution.  It should be 
noted, however, that at low rotation rates the distribution of photometric 
equatorial rotation velocities, $v$, is not the same as that derived from 
$v \sin\,i$ after deconvolution.  This is due to the fact that rotational 
modulation cannot be observed at low inclination angles, although $v\sin\,i$ 
can still be measured.  A deficiency of stars at low values of $v$ is expected 
relative to that obtained from $v\sin\,i$ measurements.  The deficiency is 
increased by the fact that rotation frequencies smaller than about 
0.1\,d$^{-1}$ probably escape detection owing to increased periodogram noise at
low frequencies.  At high rotation rates, the stellar radius is larger than the
mean radius used to determine $v$.  It is probable that this affects $v\sin\,i$
in a different way than it does $v$.  The question of differential rotation is 
also likely to affect $v$ and $v\sin\,i$ differently.  

Obtaining a distribution requires a large number of stars in order to
provide sufficient rotation rate resolution.  Fortunately, it is not really 
necessary to use the same stars in comparing the distributions of $v$ and 
$v\sin\,i$.  It is reasonable to assume that any sample of main sequence stars 
within a given temperature range will have the same distribution of 
$v \sin\,i$.  Thus one can obtain the $v$ distribution for the 1035 stars with 
$8000 < T_{\rm eff} < 10000$\,K and compare it with the $v\sin\,i$ distribution 
of an entirely different, much larger, set of main sequence stars within the 
same $T_{\rm eff}$ range.  Such a test using 875 {\it Kepler}  A  stars
in an unrestricted temperature range was made by \citet{Balona2013c}.

\begin{figure}
\centering
\includegraphics[]{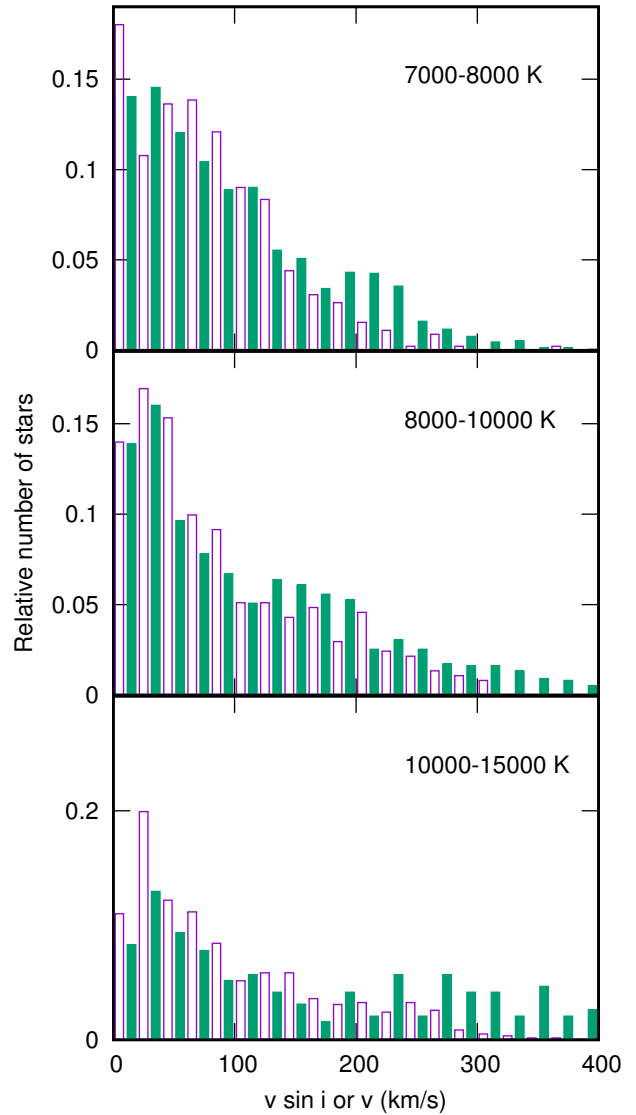}
\caption{Distribution of projected rotation velocities, $v\sin\,i$ (empty
boxes) and equatorial rotation velocities (filled boxes) for stars in
different ranges of effective temperature (labelled in degrees K).}
\label{dist}
\end{figure}

In Fig.\,\ref{dist} the $v$ distributions for stars in three temperature
ranges are shown.  Also shown are the distributions of $v\sin\,i$ for stars in
the same temperature ranges.  There are  too few stars hotter than 15000\,K 
for a meaningful analysis.  Although the two distributions are not directly
comparable due to the $\sin i$ factor, there is a clear similarity between 
them.  The tail at rapid rotation rates the number of stars in the $v$
distribution is larger than in the $v\sin i$ distribution.  This is expected 
because at large values of $v$, only a few stars with inclinations very close 
to equator-on will have large $v\sin\,i$.  

One could convolve the $v$ distribution if one knew the distribution of $i$ 
for the stars measured photometrically.  This is not known because we do not 
know the distribution of sizes and locations of the starspots.  As a test, one 
could simply assume a random distribution of $i$ for $i > 30^\circ$ with zero 
for smaller values of $i$.  The resulting convolution reduces the numbers of 
stars in the tails of the $v$ distributions, producing good agreement with the 
tails of the $v\sin i$ distributions.  This does not prove that rotational
modulation is involved, but shows that it is not difficult to obtain
agreement even in this simple case.  Once again, no rigorous statistical test 
can be made until we have a better knowledge of the underlying distribution of 
starspot sizes and locations.

The point that is being illustrated in Fig.\,\ref{dist} is that the
photometric $v$ measurements give approximately the same rotational velocity
distribution as $v\sin\,i$ measurements.  This can only be the case if $v$
measures the equatorial rotation velocity or something very close to it,
which is a different test from that shown in Fig.\,\ref{vsini} and involves
many more stars.  The implication is that whatever is responsible for the 
periodic light variation is indistinguishable from rotation.  It is found that 
about 20--40\,percent of A and B stars are presumed rotational variables.  This
fraction increases to 40--60\,percent for cooler stars.  Thus rotational light 
modulation, or some other effect indistinguishable from rotational modulation, 
is very common among all main sequence stars.

\section{Discussion}

While these results show that a co-rotating feature appears to be present in 
many hot stars, it is possible that it may not be the same as a sunspot and may 
not involve a magnetic field.  Apart from rotation, there are only two other
possibilities which might account for the observations: binarity and
pulsation.

Since the observed periods are relatively short, stars in a binary system will 
be rather close, making eclipses very probable.  One would thus expect much
higher amplitudes typical of eclipses, not the very low amplitudes of around
100\,ppm that are actually observed.  Grazing eclipses can only happen in a 
narrow range of inclination, which is inconsistent with the large fraction of 
stars observed to vary.   Furthermore, the light amplitudes often vary with
time, as expected from starspots, but not from eclipsing binaries.  Finally, 
there is the obvious observation that the periods closely agree with the
rotation periods and not the orbital periods.

Some form of standing wave might also be considered.  Indeed, this idea has 
been proposed to account for the ROTD feature in the periodogram discussed
above. In about 15\,percent of {\it Kepler} A stars, the periodogram shows a 
broad hump with a closely-spaced sharp peak at slightly higher frequency
\citep{Balona2013c}.  It was originally suggested that the broad hump may be 
due to starspots in differential rotation and the sharp peak a result of a 
reflection effect from a planet in a synchronous orbit \citep{Balona2014b}.  
Recently, it has been proposed that the broad peak may be due to Rossby waves, 
while the sharp peak is due to rotation \citep{Saio2018a}.  Rossby waves are a 
subset of inertial waves in a rotating star.  It is unlikely that the observed 
variation is due to Rossby waves alone.  In the first place, the sharp peak is 
identified as the rotational peak, so a starspot is still required in this 
explanation. Secondly, the majority of periodogram peaks are sharp, while 
Rossby waves are multiple modes which are expected to lead to broad peaks.

An explanation is required for the low frequencies in A and B stars.  If
these are not attributed to rotational modulation, it is essential to find
an alternative hypothesis.  The hypotheses discussed above do not
account for the observations and it is therefore reasonable to adopt the
rotational modulation idea until a better solution is proposed, even if the
consequences are disruptive to current thinking. 

Some A stars also appear to flare \citep{Balona2012c,Balona2013c,Balona2015a}. 
Flares might be expected in any star with spots, and this might be taken as
an indication that the spots on A and B stars are similar to those on the
Sun. The question of flares in A stars has been disputed on the grounds that 
the flaring A stars are spectroscopic binaries or that their light curves
are contaminated by fainter stars in the same aperture \citep{Pedersen2017}.  
This could well be true, but the origin of a flare in a multiple system 
cannot be determined in this way.  This can only be done if the stellar 
disks in the system are resolved. Furthermore, it is insufficient to 
attribute the flare to one or more contaminating stars.  At the very least, 
it needs to be shown that these stars are of the kind known to flare and 
that the flare could be sufficiently energetic to be visible in the glare of 
the A star.  It turns out that flares on A stars, which should be called
``superflares'' in accordance with current usage, attain energies never seen 
in cool flare stars \citep{Balona2013c, Balona2015a}. 

The problems regarding the $\delta$~Scuti stars mentioned above remain 
unresolved using current ideas of stars with radiative envelopes.  If it is
assumed that surface convection occurs in all main sequence stars, the
presence of starspots in A and B stars is no longer problematical.  A thin 
convective envelope may provide an additional mechanism for pulsational 
instability which may assist in understanding the low frequencies in 
$\delta$~Scuti stars.

It has been suggested that magnetic fields produced in subsurface convection 
zones could appear on the surface \citep{Cantiello2009,Cantiello2011}.  As in the
Sun, strong localized magnetic fields of opposite polarity lead to a weak 
global field which would be difficult to detect.  Magnetic spots with sizes 
comparable to the local pressure scale height are predicted to manifest 
themselves as hot, bright spots.  Another possibility is that differential 
rotation in the A and B stars may be sufficient to create a local magnetic 
field via dynamo action \citep{Spruit1999,Spruit2002, Maeder2004}.  

The observations presented here indicate a need for a revision of current
understanding of the outer layers of hot stars with radiative envelopes.
The ideas discussed above, as well as other possibilities of inducing surface
convection in hot stars, should be further explored.

\section*{Acknowledgments}

LAB wishes to thank the National Research Foundation of South Africa for 
financial support. 

This paper includes data collected by the {\it TESS} mission. Funding for the 
{\it TESS} mission is provided by the NASA Explorer Program. Funding for the 
{\it TESS} Asteroseismic Science Operations Centre is provided by the Danish 
National Research Foundation (Grant agreement no.: DNRF106), ESA PRODEX
(PEA 4000119301) and Stellar Astrophysics Centre (SAC) at Aarhus University. 
We thank the {\it TESS} and TASC/TASOC teams for their support of the present
work.

This work has made use of data from the European Space Agency (ESA) mission 
Gaia (\url{https://www.cosmos.esa.int/gaia}), processed by the Gaia Data 
Processing and Analysis Consortium (DPAC,\\
\url{https://www.cosmos.esa.int/web/gaia/dpac/consortium}).\\ 
Funding for the DPAC has been provided by national institutions, in particular 
the institutions participating in the Gaia Multilateral Agreement.  

This research has made use of the SIMBAD database, operated at CDS, 
Strasbourg, France.

The data presented in this paper were obtained from the Mikulski Archive for 
Space Telescopes (MAST).  STScI is operated by the Association of Universities
for Research in Astronomy, Inc., under NASA contract NAS5-2655.

\bibliographystyle{mn2e}
\bibliography{rotat}

\label{lastpage}

\end{document}